\documentclass{aa501} 
\usepackage{epsfig,times}

\makeatletter
\def\@cite#1#2{(#1\if@tempswa , #2\fi)}
\def\@citex[#1]#2{\if@filesw\immediate\write\@auxout{\string\citation{#2}}\fi
  \def\@citea{}\@cite{\@for\@citeb:=#2\do
    {\@citea\def\@citea{;\penalty\@m\ }\@ifundefined
       {b@\@citeb}{{\bf ?}\@warning
       {Citation `\@citeb' on page \thepage \space undefined}}%
\hbox{\csname b@\@citeb\endcsname}}}{#1}}
\makeatother

\title{A change in the outburst recurrence time \\ of the Rapid Burster}

\author{Nicola Masetti}

\institute{Istituto Tecnologie e Studio della Radiazione Extraterrestre,
CNR, via Gobetti 101, I-40129 Bologna, Italy}

\offprints{Nicola Masetti, masetti@tesre.bo.cnr.it}
\date{Received 2001 October 29; Accepted 2001 November 20}

\begin{document}

\abstract{An apparent sudden change in the outburst recurrence time
displayed by the Rapid Burster occurred around the end of
1999. The time between consecutive outbursts shrinked from $\approx$200
to $\approx$100 days somewhere between November 1999 and March 2000.
In parallel, the average peak intensity also decreased of a factor $\sim$2
in all outbursts occurred after November 1999 with respect to those
occurred before this date. I discuss these
results by comparing them with similar cases of changes in the recurrence
time of other transient X--ray binaries and with the behaviour of Dwarf
Novae. A viable explanation is connected with changes in the value of the
quiescence mass transfer rate from the secondary, although the detailed
origin of these changes remains at best speculative.
Further, this result has important implications for programming
time-constrained observations of the Rapid Burster both in outburst and in
quiescence.
}

\maketitle
\markboth{N. Masetti: A change in the RB outburst recurrence time}{N.
Masetti: A change in the RB outburst recurrence time}

\section{Introduction}

The Rapid Burster (MXB\thinspace1730--335; hereafter RB) is a well  
known and extensively studied low mass X--ray binary (LMXRB; see the
review by Lewin et al. 1995) located in the globular cluster
Liller 1, at a distance of $\sim$8~kpc from earth (Ortolani et al. 1996).
The RB is unique among LMXRBs in that it displays two
different kinds of X--ray bursts: Type I bursts, which are typical of
many LMXRBs harbouring a low-magnetic field neutron star (NS), and Type
II bursts.
The former events are interpreted as due to nuclear burning of accreted  
material on the surface of the NS. Type II bursts, instead, result from
spasmodic accretion onto the NS surface (see e.g. Lewin et al. 1995). Type
I and II burst emission patterns as observed with {\it RXTE} during
several RB outbursts were thoroughly studied by Guerriero et al. (1999).
Also, X--ray spectral analysis was carried out by Guerriero et al. (1999)
with {\it RXTE} during maximum and by Masetti et al. (2000) with {\it
BeppoSAX} during the decay phase.
Recently, a transient radio counterpart of the RB was observed by Moore et
al. (2000) while the system was undergoing X--ray active phases, while
the optical and infrared counterparts are still unknown or uncertain
despite the high-accuracy position of the RB determined with {\it Chandra}
(Homer et al. 2001). 

Another characteristic feature of the RB is that this source is a
recurrent transient with outbursts lasting a few weeks, followed by
quiescent, or ``off"-state, intervals which generally are $\sim$6-8 months
long (Lewin et al. 1993; Guerriero et al. 1999); thus far, no quantitative
explanation for this behaviour has still been developed, although it may
be possibly due to some accretion energy storage mechanism in the disk
around the accreting NS as described by the disk-instability model (e.g.
Lasota 2001). However, in this {\it Letter} I will show, by means of the
data collected up to now by the All-Sky Monitor (ASM) onboard {\it RXTE},
that recently this recurrence time has apparently undergone a substantial
change. Section 2 will deal with the analysis of the ASM data, and in
Section 3 a discussion of the results is given.

\section{ASM light curve data analysis}

The {\it RXTE} satellite (Bradt et al. 1993) carries an
ASM\footnote{ASM light curves are available at:\\ 
{\tt http://xte.mit.edu/ASM\_lc.html}} (Levine et al. 1996) which regularly
scans the X--ray sky in the 2--12 keV range. During each scan, data from
90-seconds long dwells are obtained for each known X--ray source covered
by the ASM. From these individual dwells (typically 5-10) a one-day
averaged light curve of the monitored X--ray sources with a daily
sensitivity of 5 mCrab is automatically computed and is available also.
Figure 1 reports the complete up-to-date (November 2001) 2--12 keV ASM
one-day averaged light curve of the RB starting on January 1996. 

\begin{figure*}
\epsfig{figure=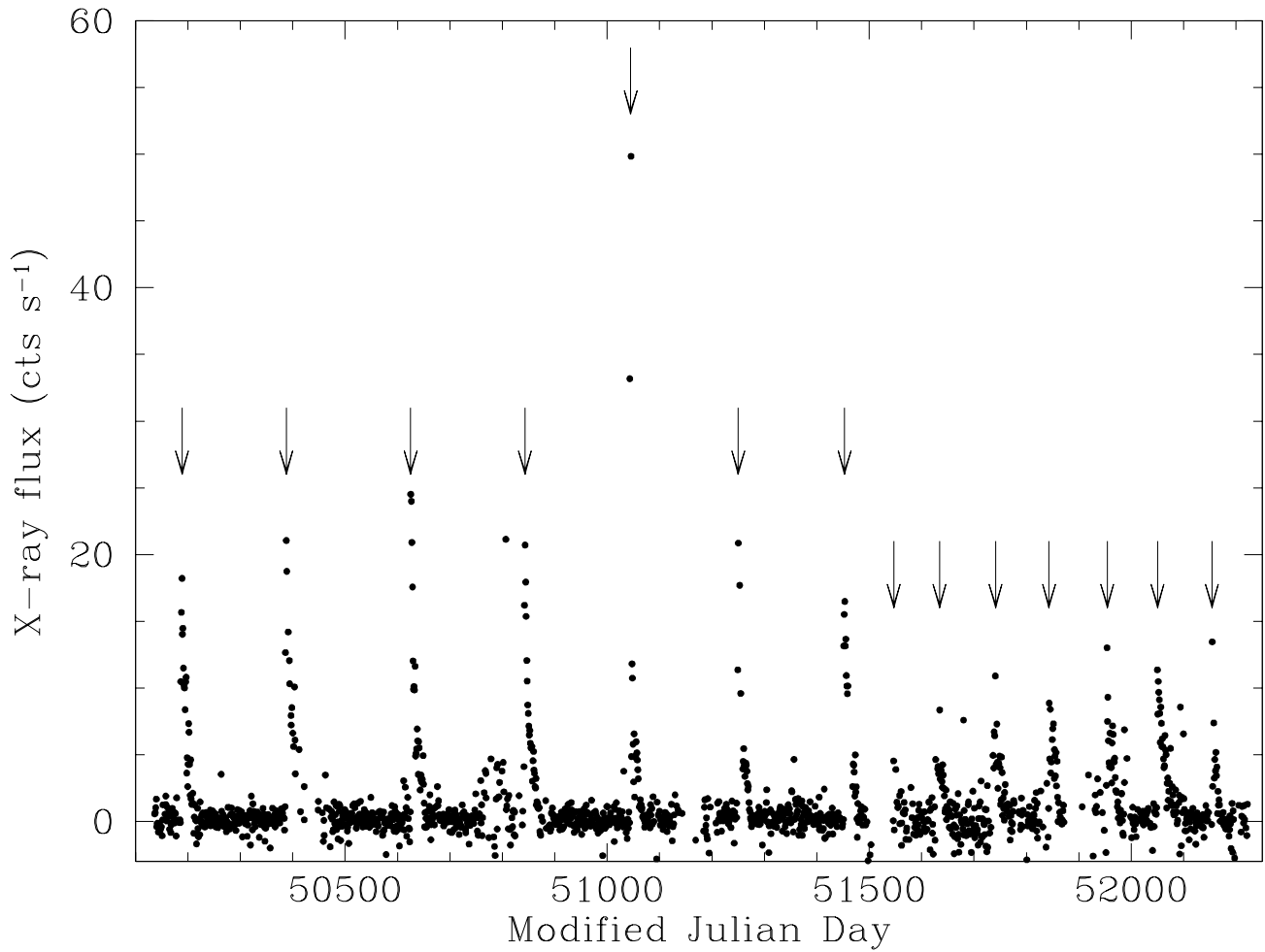,width=13.5cm,angle=270}
\caption[]{2--12 keV one-day averaged {\it RXTE} ASM light curve of the
RB. 1 ASM count s$^{-1}$ roughly corresponds to 13 mCrab assuming
a Crab-like spectrum. 
The reduced time interval between consecutive outbursts and the
different average outburst peak intensities are apparent by comparing the
data before and after MJD 51500 (i.e. 1999 November 17).
The arrows indicate the times of outburst peaks.
The errors associated with the data were not plotted in order
to have a clearer view of the overall light curve shape}
\end{figure*}

The light curve is divided into two panels for the sake of clarity in
order to show an apparent change of recurrence time in the RB outbursts
after MJD 51500 (November 17, 1999). It can also be noted that the
outburst peak intensities are significantly lower after that date, and
that the first two outbursts after that date, namely the one around MJD
51625 (March 21, 2000) and the previous, albeit less well determined, one
around MJD 51550 (January 6, 2000), have the most scattered shape and the
lowest peak intensity among the whole ASM set shown in Fig. 1. 

To better study and to quantify this phenomenon, I performed a timing
analysis of the entire {\it RXTE} ASM daily averaged data set using a
Discrete Fourier Transform (DFT) algorithm (for a thorough description see
Press et al. 1992).
The Power Spectral Density (PSD) obtained with the DFT
algorithm and corresponding to this data set is shown in Fig. 2.
As one can see, the two most prominent
peaks (marked by the arrows) fall at 0.48$\times$10$^{-2}$ and
0.97$\times$10$^{-2}$ cycles d$^{-1}$, which correspond to $\sim$210 and
$\sim$100 days, respectively.

\begin{figure}
\begin{center}
\epsfig{figure=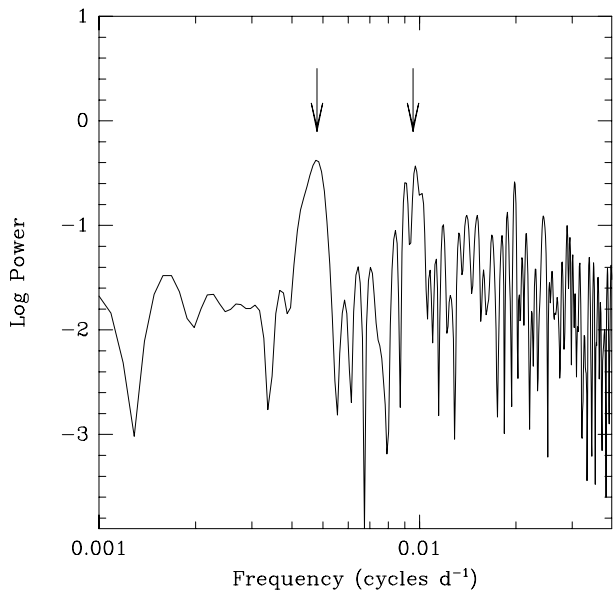,width=10cm}
\end{center}
\vspace{-1.5cm}
\caption[]{PSD of the entire one-day averaged {\it RXTE} ASM light curve
of the RB shown in Fig. 1. The two most prominent periodicities at
$\sim$210 and $\sim$100 days are indicated with the arrows}
\end{figure}

In order to get a better insight on this result, and most of all to
prove that the shorter recurrence periodicity of Fig. 2 is not simply an
alias of the main one produced by the Fourier analysis of the ASM data
set, I also ran the DFT algorithm on the data collected after MJD 51500.
The corresponding PSD is reported in Fig. 3. From this figure it appears
evident that only the peak corresponding to the higher frequency, here at
0.95$\times$10$^{-2}$ cycles d$^{-1}$ ($\approx$105 days), has survived.
Instead, at the position of the peak associated with the lower frequency
(again marked with an arrow) there is no significant signal from the
considered subset of data points.

Thus, the $\sim$6-8 months long outburst recurrence time seems to have
suddenly halved about two years ago and is still following this
periodicity since then. The same can be said concerning the outburst
average peak intensities, which changed from $\sim$25 counts s$^{-1}$
before November 1999 to $\sim$12 counts s$^{-1}$ after that date. 
Identical results are obtained if the DFT algorithm is applied to the
dwell-by-dwell ASM light curve of the RB. 

Furthermore, to check the reliability of this result for the sake of 
completeness, I also inspected the periodicities displayed by the ASM
data by running a non-DFT period searching algorithm based on the 
Phase Dispersion Minimization (PDM) method (Stellingwerf 1978).
The results obtained with the PDM algorithm from the analysis of both 
the daily averaged and the dwell-by-dwell ASM data points do not show 
any significant difference with respect to those coming from the 
application of the DFT method.

\begin{figure}
\begin{center}
\epsfig{figure=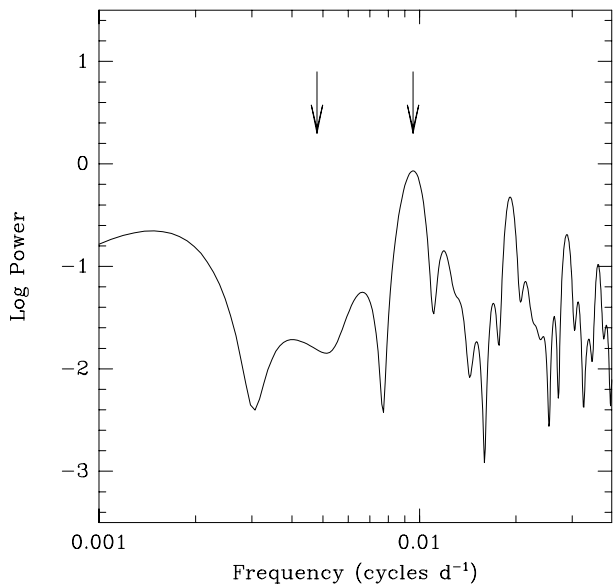,width=10cm}
\end{center}
\vspace{-1.5cm}
\caption[]{The same as Fig. 2, but computed on the one-day averaged {\it
RXTE} ASM data after MJD 51500. Only the shorter periodicity of Fig. 2 is
present here, while no significant signal is detected at the position of
the longer one}
\end{figure}

\section{Discussion}

I have shown, from the analysis of the long-term monitoring of the RB by
means of the {\it RXTE} ASM, that a change in the recurrence time of this
source suddenly occurred around the end of 1999. In parallel, a 
decrease in the outburst peak intensity is seen occurring at the same
time. Possibly, the first outburst (around MJD 51550) after this sudden
change was aborted or at least not fully developed. Since then, the RB
outbursts occurred following this new recurrence time. 

Exploring the causes of this behaviour is not an easy task. As a first
attempt, one can consider the possibility that the recurrence time
reflects the orbital period of the system and that its change is connected
with a modification in the orbit of the system. This is however
unacceptable for several reasons. First, given that the RB is a LMXRB 
clearly hosting a NS (as it displays X--ray bursts) which is accreting 
via Roche lobe overflow, an orbital period of $\sim$200 (or even of 
$\sim$100) days is far too long to be reconciled with this scenario. 
Second, a reduction of 50\% in the orbital period over
a time comparable with the period itself was never seen in LMXRBs and very
likely in any other type of close binary, and in any case it would
indicate a severe orbital instability phase eventually leading to a very
quick disruption of the system; no indication of all this is however
apparent from the ASM data set shown in Fig. 1. Third, a long-period
binary system would hardly survive the interactions occurring within the
core of a globular cluster. Fourth, optical and near-infrared observations 
of the RB error box as determined by {\it Chandra} do not reveal the presence
of any remarkable object with anomalous colors and/or resembling a giant 
(Homer et al. 2001). Therefore, the eventuality of a connection of
the decrease in the recurrence time with a change in the supposed (long) 
orbital period of this system can reasonably be excluded. 

An alternative, and in my opinion more viable, explanation may come by
considering a sudden (altough possibly small) change in the average 
quiescence mass transfer rate $\dot{M}_{\rm q}$ from the secondary.
Indeed, according to the disk-instability model, an increase in
$\dot{M}_{\rm q}$ causes the recurrence time to shrink substantially (see
e.g. Osaki 1996).
An example of connection between these two parameters comes by considering
the behaviour of SU UMa-type and ER UMa-type systems: both are 
short-period Dwarf Novae (DNe) in which a white dwarf is accreting from a
late-type dwarf star via Roche lobe overflow, but objects belonging to the
ER UMa-type group are characterized by shorter outburst cycles and smaller
amplitude outbursts (Kato \& Kunjaya 1995). These differences are
interpreted as due to the fact that $\dot{M}_{\rm q}$ from the secondaries
in ER UMa-type systems is higher and closer to the critical value which
marks the border between stable and unstable accretion disks (Osaki 1995,
1996).

Thus one may suppose that, given its frequent outbursts, the RB is close 
to the critical quiescence mass transfer rate $\dot{M}_{\rm q,crit}$ for
(in)stability, and that an increase of $\dot{M}_{\rm q}$ may have brought
the system even closer 
to persistence; this may have therefore started the production of 
outbursts with shorter recurrence times and lower peak intensities.
It must however be noted that, in the case of the RB and more generally 
of LMXRBs, $\dot{M}_{\rm q}$ should not be confused with the total
(quiescence plus outburst) mass transfer rate $<$$\dot{M}$$>$ averaged 
over an outburst cycle, which is an indicator the total energy emitted 
during the outburst, and in turn for the outburst peak intensity: looking
at Fig. 1, one can argue that the value of this parameter has instead 
clearly reduced after MJD 51500. I also remark here that $\dot{M}$ in
transient LMXRBs increases substantially during outburst because of
strong X--ray irradiation of the secondary; this effect is instead
practically absent in DNe due to the nature of the accreting object (van
Paradijs 1996).

The trend observed in the light curve of Fig. 1 is also reminiscent of the
behaviour followed by Type II bursts displayed by the RB during outbursts.
Their time behaviour is like that of a relaxation oscillator: their 
fluence $E$ is roughly proportional to the time interval, $\Delta t$, 
to the following burst (Lewin et al. 1976) and their origin is known to 
be due to spasmodic accretion onto the NS (Lewin et al. 1993).
Moreover, the long-term behaviour of the RB X--ray light curve is also
similar to the well-known outburst pattern observed in the optical from
DNe, in the sense that the longer is the time spent in quiescence
after the last outburst, the larger will be the accretion energy stored in
the disk and thus the more intense will be the subsequent outburst (for a 
complete phenomenological description of DNe, see Warner 1995). This is a 
further indication that an accretion phenomenon is responsible for the 
observed recurrence time change.

The reasons for this increase of $\dot{M}_{\rm q}$ are however not
straightforward: in particular, it is not clear whether this is due to a
increased X--ray heating produced by the quiescent NS emission on the
outer layers of the secondary star (as described by e.g. Hameury et al. 
1986), or to a small change in the orbital parameters which shrinked the
Roche lobe of the secondary star, or to other, more complex phenomena.
It must be however stressed that a detailed analysis of the causes of this
change in the RB outbursts recurrence time is beyond the scopes of this
{\it Letter}: here I just wanted to draw the reader's attention on the
result as it is. 

This variation in the outburst recurrence time, albeit peculiar, is
actually not an
unique behaviour among transient X--ray binaries. Indeed, two recurrent
Soft X--ray Transients, namely 4U 1630$-$47 (suspected to harbour a black
hole) and Aql X-1 (known to contain a NS) have shown changes in their
outburst recurrence times in the past. Concerning 4U 1630$-$47, Kuulkers
et al. (1997) found that the quiescent period between outbursts increased
in a linear fashion from $\sim$600 to $\sim$700 days between the years
1984 and 1987. As regards Aql X-1, Kitamoto et al. (1993) noted that the
outburst recurrence time increased substantially over the years, going
from 125 days as observed between 1969 and 1979 to 309 days in the
period 1987-1992. In both cases, and in particular for Aql X-1, changes in
the value of $\dot{M}_{\rm q}$ from the secondary (possibly induced by
chromospheric activity) or instabilities in the accretion disk were
invoked to explain the observed behaviour, although none of these
modelizations seems to satisfactorily describe these apparent changes in
the ``internal clock" of these systems (e.g. Kitamoto et al. 1993). 

As a final remark, it should be noteworthy to say that the result
illustrated here can also be useful for programming future multiwavelength
observations of the next outbursts coming from the RB. Indeed, if this
new recurrence time shown here for the RB outbursts will be maintained in
the future, I expect that the next RB outburst will start around the last
decade of December 2001. Moreover, a deep X--ray observation of the RB in
quiescence performed with high-sensitivity instruments onboard
last-generation satellites such as {\it Chandra} and {\it XMM-Newton} is
highly desireable in order to provide a high precision measurement of the
quiescent flux from this system and a comparison with the only quiescent
RB detection obtained -- with {\it ASCA} -- so far (Asai et al. 1996).
Such a comparison will help to give a better insight on the proposed
inrerpretation (the increase of $\dot{M}_{\rm q}$) as the cause of the
observed change in the RB outburst recurrence time.

\begin{acknowledgements}
I would like to thank Filippo Frontera, Sergio Campana and Luigi Stella
who triggered this analysis; I also thank Eliana Palazzi and again Luigi
Stella for several important comments on the manuscript. I am grateful to
the anonymous referee for several comments which helped me to improve this
{\it Letter}. ASM data were provided by the {\it RXTE} ASM teams at MIT
and at the {\it RXTE} SOF and GOF at NASA's GSFC.
\end{acknowledgements}


\begin{thebibliography}{}

\bibitem{} Asai, K., Dotani, T., Kunieda, H., Kawai, N., 1996, PASJ, 48,
	L27

\bibitem{} Bradt, H.V., Rothschild, R.E., \& Swank, J.H., 1993, A\&AS, 97,
	355

\bibitem{} Guerriero, R., Fox, D., Kommers, J., et al., 1999, MNRAS, 307,
	179

\bibitem{} Hameury, J.M., King, A.R., Lasota, J.-P., 1986, A\&A, 162, 71

\bibitem{} Homer, L., Deutsch, E.W., Anderson, S.F., \& Margon, B., 2001,
	AJ, 122, 2627

\bibitem{} Kato, T., Kunjaya, C., 1995, PASJ, 47, 163

\bibitem{} Kitamoto, S., Tsunemi, H., Miyamoto, S., \& Roussel-Dupre,
	D., 1993, ApJ, 403, 315

\bibitem{} Kuulkers, E., Parmar, A.N., Kitamoto, S., Cominsky, L.R., \&
	Sood, R.K., 1997, MNRAS, 291, 81

\bibitem{} Lasota, J.-P., 2001, New Astron. Rev., 45, 449

\bibitem{} Levine, A.M., Bradt, H.V., Cui, W., et al., 1996, ApJ, 469, L33

\bibitem{} Lewin, W.H.G., Doty, J., Clark, G.W., et al., 1976, ApJ, 207, L95

\bibitem{} Lewin, W.H.G., van Paradijs, J., \& Taam, R.E., 1993, Space
	Sci. Rev., 62, 223

\bibitem{} Lewin, W.H.G., van Paradijs, J., \& Taam, R.E., 1995,
        in: Lewin, W.H.G., van Paradijs, J., \& van den Heuvel, E.P.J.,
	(eds.) ``X--ray Binaries", Cambridge Univ. Press, p. 175

\bibitem{} Masetti, N., Frontera, F., Stella, L., et al., 2000, A\&A, 363,
	188

\bibitem{} Moore, C.B., Rutledge, R.E., Fox, D.W., et al., 2000, ApJ, 532,
	118

\bibitem{} Ortolani, S., Bica, E., \& Barbuy, B., 1996, A\&A, 306, 134

\bibitem{} Osaki, Y., 1995, PASJ, 47, L11

\bibitem{} Osaki, Y., 1996, PASP, 118, 39

\bibitem{} Press, W.H., Teukolsky, S.A., Wetterling, W.T., \& Flannery,
	B.P., 1992, Numerical Recipes. Cambridge Univ. Press, Cambridge

\bibitem{} Stellingwerf, R.F., 1978, ApJ, 224, 953

\bibitem{} van Paradijs, J., 1996, ApJ, 464, L139

\bibitem{} Warner B., 1995, Cataclysmic Variable Stars. Cambridge Univ.
	Press, Cambridge

\end{thebibliography}
\end{document}